\documentclass[aps,prl,showpacs,twocolumn,superscriptaddress,floatfix]{revtex4}

\newcommand{\gi}{\dot{\gamma}_{\hbox{\rm\scriptsize init}}}
\newcommand{\gt}{\dot{\gamma}_{\hbox{\rm\scriptsize true}}}
\newcommand{\gti}{\dot{\gamma}_{\hbox{\rm\scriptsize true}}^\infty}
\newcommand{\ei}{\varepsilon^\infty}
\newcommand{\ga}{\dot{\gamma}_{\hbox{\rm\scriptsize appl}}}
\newcommand{\gm}{\dot{\gamma}_{\hbox{\rm\scriptsize -}}}
\newcommand{\gp}{\dot{\gamma}_{\hbox{\rm\scriptsize +}}}
\newcommand{\gl}{\dot{\gamma}_{\hbox{\rm\scriptsize 1}}}
\newcommand{\gh}{\dot{\gamma}_{\hbox{\rm\scriptsize 2}}}

\renewcommand{\ss}{\sigma^\star}

\usepackage{graphicx}
\usepackage{dcolumn}
\usepackage{epsfig}
\usepackage{latexsym}
\usepackage{amsfonts}
\usepackage{amssymb}

\begin{document}

\bibliographystyle{apsrev}
\title{Competition between shear banding and wall slip in wormlike micelles}
\author{M. Paul Lettinga}
\email{p.lettinga@fz-juelich.de}
\affiliation{IFF, Institut Weiche Materie, Forschungszentrum J\"{u}lich,D-52425
J\"{u}lich, Germany}
\author{S\'{e}bastien Manneville}
\affiliation{Laboratoire de Physique, Universit\'e de Lyon -- \'Ecole Normale Sup\'erieure de Lyon  -- CNRS UMR 5672\\46 all\'ee d'Italie, F-69364 Lyon cedex 07, France}

\date{\today}
\begin{abstract}
The interplay between shear band (SB) formation and boundary conditions (BC) is investigated in wormlike micellar systems (CPyCl--NaSal) using ultrasonic velocimetry coupled to standard rheology in Couette geometry. Time-resolved velocity profiles are recorded during transient strain-controlled experiments in smooth and sand-blasted geometries. For stick BC standard SB is observed, although depending on the degree of micellar entanglement temporal fluctuations are reported in the highly sheared band. For slip BC wall slip occurs only for shear rates larger than the start of the stress plateau. At low entanglement, SB formation is shifted by a constant $\Delta\dot{\gamma}$, while for more entangled systems SB constantly ``nucleate and melt.'' Micellar orientation gradients at the walls may account for these original features.
\end{abstract}


\maketitle

During the past two decades, shear banding (SB), i.e. the shear-induced coexistence of macroscopic bands with widely different viscosities, has been evidenced in a large range of complex fluids \cite{Reviews}. Sheared dispersions of surfactant wormlike micelles have attracted considerable attention due to their practical use in industry, but also because they challenged the physicists to address a non-equilibrium problem with concepts from thermodynamics \cite{Reviews,Grand97,Berret94}. Indeed, rheological measurements show that the flow curve of shear-banding systems, i.e. the measured shear stress $\sigma$ vs. the applied shear rate $\dot{\gamma}$, presents a plateau at a well-defined shear stress $\ss$ over a given range of shear rates \cite{Rehage91}, very similar to the plateau in pressure as a function of overall concentration of a demixed system. As for equilibrium phase transitions, it has been suggested that the flow can be either metastable or unstable for SB formation, depending on the applied shear rate~\cite{Dhont99,Olmsted99c,Grand97,Berret97}. The formation of two coexisting SB, bearing the local shear rates $\gl$ and $\gh$ that mark respectively the lower and upper limits of the stress plateau, constitutes a pathway for the relaxation of the excess stress in the initially linear flow.
Stress relaxation can, however, also occur through apparent wall slip. Slip phenomena are ubiquitous in polymers \cite{Drda95,Boukany08} and soft glassy materials \cite{glasslip}. Wall slip has also been reported in shear-thinning wormlike micellar systems~\cite{Lopez04,Becu04} but its connection with SB has been underexposed. Still, information on the interplay between wall slip and a flow instability like SB are essential for fully understanding the behavior of complex fluids.

In this Letter, wall slip is shown to compete with SB formation by offering an alternative route for stress relaxation. We use {\it tunable boundary conditions} (BC) at the walls as an experimental tool to probe the effect of wall slip on the flow behavior of cetylpyridinium chloride/sodium salicylate (CPyCl--NaSal) micellar solutions at 6 and 10~wt.~\% in 0.5~M NaCl brine at $23^\circ$C. We enforce ``stick'' BC  by using a rough sand-blasted Plexiglas Couette cell and partial ``slip'' BC by using a smooth Plexiglas cell \cite{cells}. The competition between SB formation and wall slip after shear rate quenches is addressed through simultaneous rheological \cite{rheological} and time-resolved velocity profiles measurements. For the latter, we use ultrasonic speckle velocimetry (USV) \cite{Manneville04b} since the sand-blasted cell is not transparent and optical techniques as in \cite{Salmon03,Hu05,Miller07} would be too difficult to implement. We show that with slip BC, wall slip occurs only for shear rates larger than the start of the stress plateau for both the concentrations under study. The extent to which SB formation is frustrated by wall slip strongly depends, however, on the degree of micellar entanglement. Very large temporal fluctuations are reported in the more concentrated sample for both BC.

\begin{figure}[htb]
\includegraphics[width=.50\textwidth]{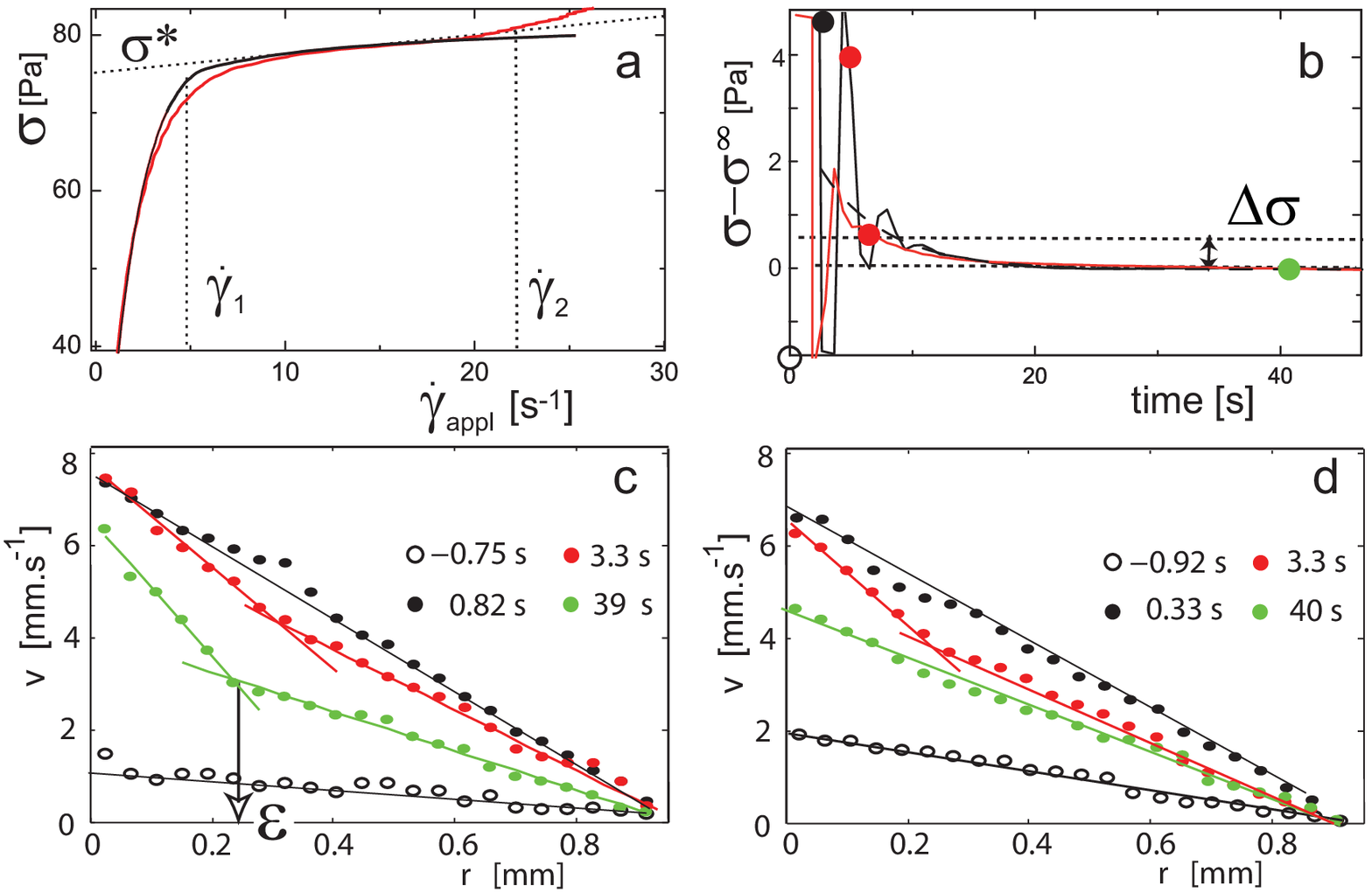}
\caption{\label{quench} (a) Steady-state flow curves of 6 wt.~\% CPyCl--NaSal for stick (black) and slip BC (red) for a shear rate sweep of 1200~s. (b) Stress responses $\sigma(t)-\sigma^{\infty}$ after a shear rate quench to $\ga=8$~s$^{-1}$ at $t=0$ for stick (black, $\gi=0.8$~s$^{-1}$) and slip BC (red, $\gi=2$~s$^{-1}$). The dashed line indicates an exponential decay with a characteristic time of 10~s. (c) Velocity profiles $v(r,t)$ for stick BC at various times during the quench shown in (b) (see colored dots in (b)). $r$ denotes the radial position from the inner rotating cylinder. (d) Same as (c) for slip BC.}
\end{figure}

The flow curve of 6~wt.~\% CPyCl--NaSal shown in Fig.~\ref{quench}(a) for stick and slip BC reveals a stress plateau at $\sigma^\star\simeq 75$~Pa that extends from $\gl\simeq 4.5$ to $\gh\simeq 22$~s$^{-1}$, with a slight tilt due to the curvature of the Couette cell \cite{Salmon03}. Interestingly the flow curve for slip BC does not show such a sharp bend at $\gl$ as with stick BC. Fig.~\ref{quench}(b) presents the stress responses for quenches from $\gi$ located in the low shear regime to $\ga=8$~s$^{-1}$ located in the {\it beginning} of the stress plateau. Conform to earlier experiments \cite{Grand97,Berret97}, the stress shows a slow decay after an initial overshoot and a few oscillations. As in Ref.~\cite{Berret97}, we define the amplitude of this slow relaxation as the {\it excess} stress $\Delta\sigma=\sigma^M-\sigma^\infty$, where $\sigma^M$ is the ``mechanical'' stress at the end of the oscillations and $\sigma^\infty$ is the steady-state shear stress. Although the initial overshoot is more pronounced for stick BC, the stress responses for $t\gtrsim 10$~s are very similar for both BC. Yet, depending on the BC, velocity profiles display radically different behaviors that persist in the steady state. As seen from Fig.~\ref{quench}(c) and (d), linear profiles are recorded just before and after the shear rate quench for both BC. For stick BC, a high SB develops within a few seconds at $\varepsilon=\delta/e\simeq 0.5$, where $\delta$ is the width of the SB and $e$ the gap width. The interface then migrates towards its final position in agreement with previous observations \cite{Hu05}. For slip BC, however, SB do not fully develop (see the velocity profile at $t=3.3$~s in Fig.~\ref{quench}(d)) and the system rather slips to reach a steady state characterized by a {\it homogeneous} shear flow with substantial wall slip (about 40~\%) at the inner cylinder, as in a recent report on DNA dispersions~\cite{Boukany08}.

\begin{figure}[htb]
\includegraphics[width=.4\textwidth]{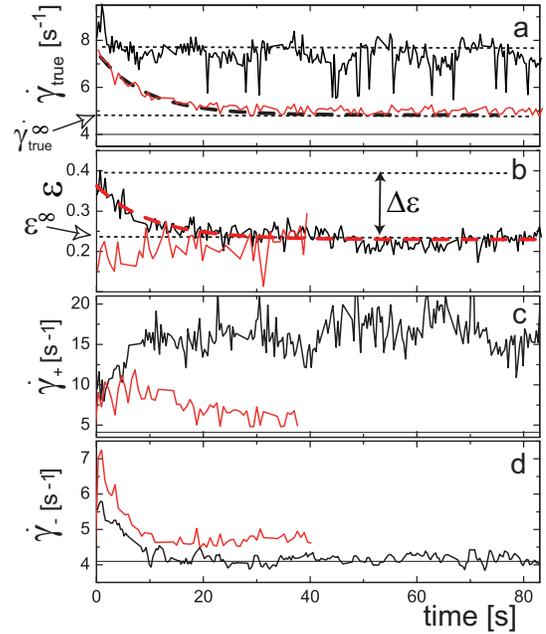}
\caption{\label{response} Analysis of the velocity data corresponding to the shear rate quenches to $8$~s$^{-1}$ with stick (black) and slip BC (red). (a) True shear rate $\gt(t)$. (b) Proportion of the high SB $\varepsilon(t)$. (c) Shear rate in the high SB $\gp(t)$ and (d) in the low SB $\gm(t)$. The dashed lines indicate exponential decays with the same time constant of 10~s.}
\end{figure}

Figure \ref{response} provides the analysis of the time-resolved velocity measurements after quenches to $\ga=8$~s$^{-1}$, as in Fig.~\ref{quench} (c) and (d). Each velocity profile was analyzed to extract the true shear rate $\gt(t)$, the proportion of highly sheared material $\varepsilon(t)$, and the local shear rates $\dot{\gamma}_{\pm}(t)$ in each SB \cite{defs}. As noted above our results for {\it stick} BC are consistent with previous data where no significant wall slip was reported \cite{Hu05,Salmon03}. They also reveal two important new features: (i) the presence of noticeable fluctuations in both $\gt(t)$ and $\gp(t)$ while $\varepsilon(t)$ and $\gm(t)$ remain roughly constant for $t\gtrsim 30$~s and (ii) the fact that the position of the SB settles with the same dynamics as the shear stress. For {\it slip} BC, Fig.~\ref{response}(a) shows that (i) the imposed shear rate $\ga$ cannot be sustained although initially $\gt\simeq\ga$ and (ii) wall slip sets in immediately and has the same time constant as the stress relaxation since $\gt(t)$ and $\sigma(t)$ follow the same decay. Figs.~\ref{response}(b) and (c) reveal that shear banding is observed during the build up of wall slip. A high SB is formed with $\varepsilon\simeq 0.2$ and $\gp\simeq 10$~s$^{-1}$, a value close to the initial shear rate for stick BC. However, for slip BC, $\gp$ rapidly drops and approaches $\gm$, which leads to the loss of banding structure (hence the lack of $\varepsilon(t)$ and $\dot{\gamma}_\pm$ data for $t\gtrsim 40$~s) and to linear profiles with $\gt\simeq 5$~s$^{-1}$ in the steady state. We conclude that at $\ga=8$~s$^{-1}$ the excess stress relaxes fully due to wall slip for slip BC, while for stick BC it relaxes by SB formation. Both processes have the same time constants since the decay of $\gt(t)$ for slip BC is the same as the settling of the SB through $\varepsilon(t)$ for stick BC. As a consequence the stress relaxations for stick and slip BC are also similar (see the dashed line in Fig.~\ref{quench}(b)).

\begin{figure}[htb]
\includegraphics[width=.4\textwidth]{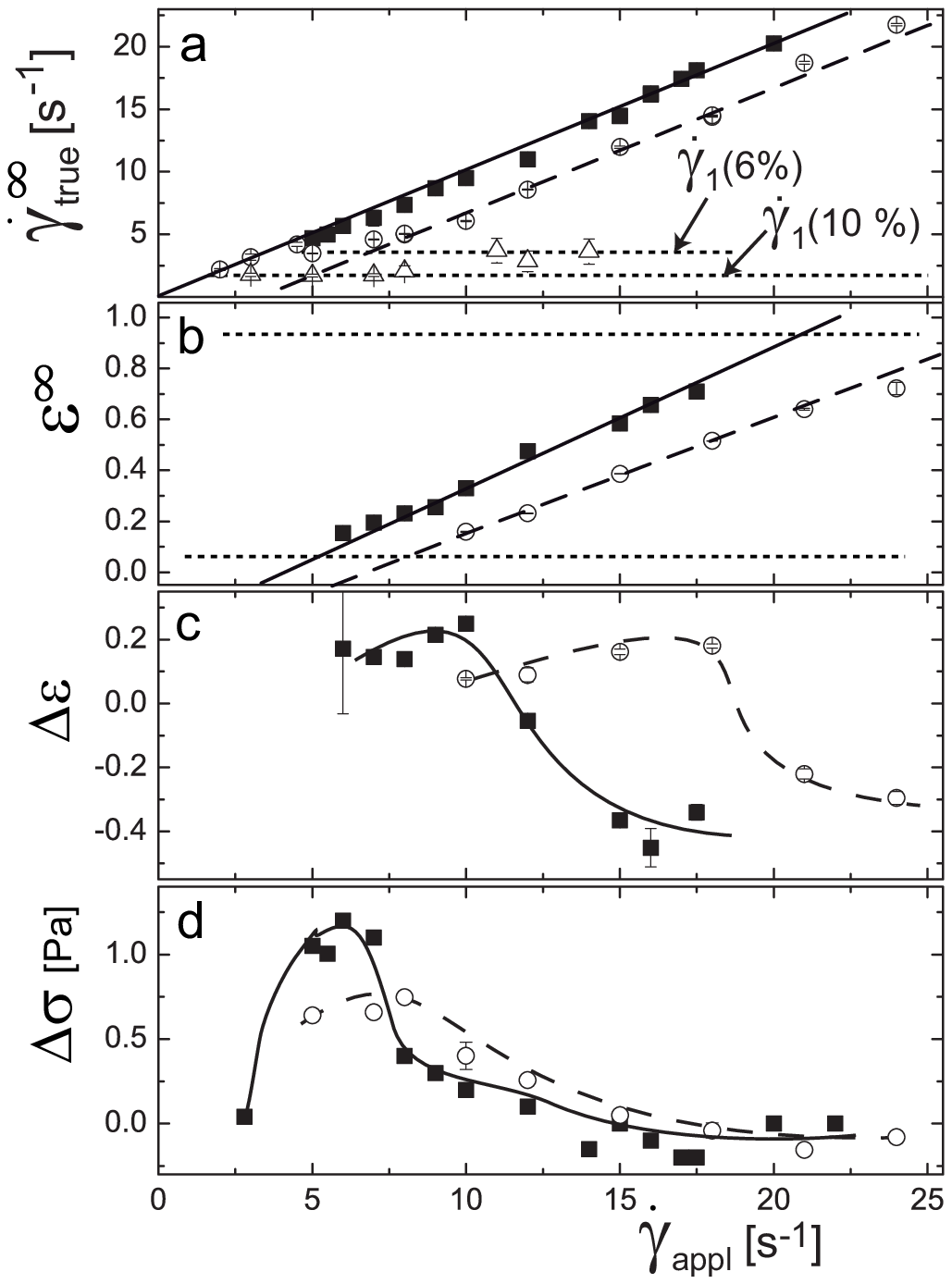}
\caption{\label{truerate}Steady state values of (a) the true shear rate $\gti$ and (b) the proportion of the high SB $\ei$, and amplitudes of the relaxations of (c) the proportion of the high SB $\Delta\varepsilon$ and (d) the excess shear stress $\Delta\sigma$ as a function of the imposed shear rate $\ga$. Solid (open) symbols refer to stick (slip) BC. All data are for 6~wt.~\% CPyCl--NaSal except for the triangles in (a) which are for 10~wt.~\% CPyCl--NaSal with slip BC. In (a) the solid line is $\gt=\ga$, while the dashed line is $\gt=\ga-3.4$~s$^{-1}$ and the dotted lines indicate $\gl$ for 6 and 10~wt.~\% CPyCl--NaSal. The dotted lines in (b) show the resolution limit for SB detection \cite{defs}. The lines in (c) and (d) are to guide the eye.}
\end{figure}

Quenches were repeated as described above for final shear rates $\ga$ covering almost the whole stress plateau \cite{range}. Fig.~\ref{truerate} presents the steady state values of the true shear rate $\gti$ and the proportion of the high SB $\ei$, as well as the amplitude of the relaxation of $\varepsilon(t)$ (noted $\Delta\varepsilon$ and defined in Fig.~\ref{response}(b)) and that of the stress relaxation $\Delta\sigma$. As shown by the solid line $\gt=\ga$ in Fig.~\ref{truerate}(a), stick BC apply for the sand-blasted cell. Moreover the linear behavior of $\ei$ vs. $\ga$ is consistent with the ``lever rule'': $\varepsilon^{\infty}=(\ga-\gl)/(\gh-\gl)$ (see solid line in Fig.~\ref{truerate}(b)) with $\gl=3.4\pm 0.2$~s$^{-1}$ and $\gh=22.4\pm 0.5$~s$^{-1}$ in satisfactory agreement with both the flow curve and the steady state values of the local shear rates $\gm=4.3\pm 0.3$~s$^{-1}$ and $\gp=22\pm 1$~s$^{-1}$ measured from the velocity profiles \cite{defs}. These observations not only confirm previous results in the absence of wall slip \cite{Hu05,Salmon03,Miller07} but also allow us to evidence the migration of the high SB towards the stator (i.e. $\Delta\varepsilon<0$) for deep quenches.
For slip BC, $\gt=\ga$ only holds when $\ga<\gl$. Wall slip is observed over the whole stress plateau and $\gti$ is shifted by a constant $\Delta\dot{\gamma}\simeq 3.4$~s$^{-1}$ with respect to stick BC. If SB occurs in the presence of wall slip, then one expects $\ei$ to be shifted by the same amount. Figure~\ref{truerate}(b) shows that SB indeed sets in for $\ga>\gl+\Delta\dot{\gamma}\simeq 8$~s$^{-1}$. However the slope of $\ei$ vs $\ga$ is slightly smaller than for stick BC leading to a shift that increases with $\ga$ (see dashed line in Fig.~\ref{truerate}(b)). The same observation holds for Fig.~\ref{truerate}(c) where the shift between the $\Delta\varepsilon$ curves is seen to increase up to about 15~s$^{-1}$ for the highest achievable $\ga$. This suggests a more subtle influence of wall slip on SB than a mere shift due to the difference between $\ga$ and $\gt$ but remains questionable due to surface instability for very deep quenches.
Finally, if one assumes that the viscosity of the slip layer at the rotor does not depend on $\ga$ throughout the stress plateau, then a constant $\Delta\dot{\gamma}$ corresponds to some {\it constant stress} released by wall slip. Figure~\ref{truerate}(d) shows that the excess stress $\Delta\sigma$ is most affected by the BC for $\ga=5$--8~s$^{-1}$ (where $\Delta\sigma$ is about twice smaller for slip BC than for stick). At larger $\ga$, $\Delta\sigma$ follows roughly the same decay for both BC.

\begin{figure}[htb]
\includegraphics[width=.4\textwidth]{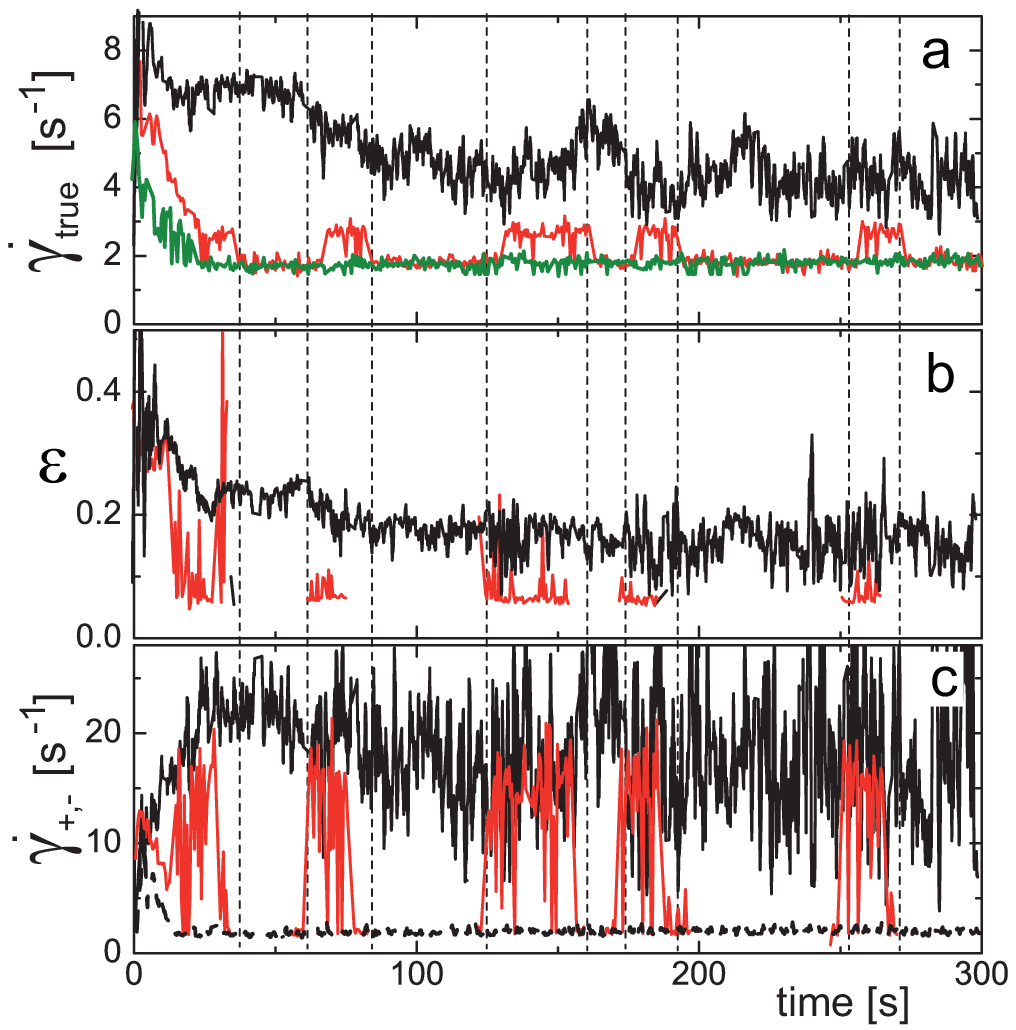}
\caption{\label{Fig4}Responses to shear rate quenches from $\gi=0.8$~s$^{-1}$ to $\ga=8$~s$^{-1}$ for stick (black) and slip BC (red) in 10~wt.~\% CPyCl--NaSal. (a) True shear rate $\gt(t)$. The green line is the response to a quench to $\ga=5$~s$^{-1}$ for slip BC. (b) Proportion of the high SB $\varepsilon(t)$. (c) Shear rate in the high SB $\gp(t)$ (solid) and low SB $\gm(t)$ (dashed, only stick BC). The vertical dashed lines indicate the times where SB nucleate and melt (see text). The gaps in the slip data in (b) and (c) are due to the lack of SB \cite{defs}.}
\end{figure}

It is interesting to see how the balance between SB and wall slip changes for a more entangled system, e.g. a 10~wt.~\% CPyCl--NaSal sample, as studied by L\'{o}pez-Gonz\'{a}lez {\it et al.} \cite{Lopez04}. The response to shear rate quenches for this system, where $\sigma^\star\simeq 158$~Pa, $\gl\simeq 1.7$~s$^{-1}$ and $\gh\simeq 20$~s$^{-1}$, is plotted in Fig.~\ref{Fig4}. Even in the sand-blasted cell where stick BC are supposed to be valid, the true shear rate never coincides with $\ga$. Moreover considerable fluctuations in $\gp$ are observed, while $\gm$ remains much smoother. For slip BC and $\ga=5$~s$^{-1}$, we observe that the sample slips to the shear rate $\gl$ at the start of the stress plateau where there is no excess stress (see the green line in Fig.~\ref{Fig4}(a)). A quench to a higher shear rate of $\ga=8$~s$^{-1}$ reveals that $\gt$ jumps from $\gl$ to significantly higher values over short time windows indicated by vertical dashed lines in Fig.~\ref{Fig4}. Velocity profiles also show that, when $\gt>\gl$, a small but detectable high SB forms with $\varepsilon\gtrsim 0.1$ and $\gp\simeq 15$--20~s$^{-1}$. In other words the high-shear state is formed over short periods of time and is unstable over longer times, which is reminiscent of ``nucleation and melt'' events typical of metastability. Since $\gt\approx\gl$ up to $\ga\simeq 12$~s$^{-1}$ (see open triangles Fig.~\ref{truerate}(a)), wall slip dominates SB formation over the full accessible part of the stress plateau.

In summary, we established that under slip BC wall slip is observed only for shear rates larger than the start of the stress plateau, i.e. $\dot{\gamma}>\gl$. For the 6~wt.~\% sample, SB formation is suppressed by wall slip over the first $\Delta\dot{\gamma}\simeq 3.4$~s$^{-1}$ into the stress plateau, while for higher shear rates stable SB are observed together with partial slip. For 10~wt.~\% CPyCl--NaSal, ``nucleation and melt'' is observed over the full accessible part of the stress plateau. This has the important implication that for both concentrations wall slip acts as to stabilize the bulk flow.

By combining presently available theories we may interpret our results along the following line of argumentation. Strong gradients in the shear rate can build up at the wall, assuming that wormlike micelles preferentially align with the smooth walls, i.e. that gradients in orientation are intrinsically present at the wall, see Ref.~\cite{sliptheory}. Orientation gradients grow when the system is quenched into the plateau region, see Refs.~\cite{Dhont99,Olmsted99c}. The stress that is stored in the system after the quench needs to diffuse in order for the system to relax, see Ref.~\cite{Radulescu03}. These latter two processes should be independent of whether the gradients are present in the bulk or at the wall. Combining these arguments one can explain the observation that no apparent wall slip is observed below the stress plateau because in this region the flow is stable and gradients at the walls or in bulk do not grow. It also follows that the time constants of $\gt$ for slip BC and $\varepsilon$ for stick BC are comparable in Fig.~\ref{response}, resulting in similar stress decays (see Fig. \ref{quench}b), since the same stress diffusion is needed in both cases. Once the SB have settled $\gm=\gl$ holds for the low SB both for stick and slip BC. Fluctuations between both conditions can now easily occur since no stress diffusion is needed. This may account for the difference between the stable low SB and the fluctuating high SB (see Fig.~\ref{Fig4}(c)), and for the fast formation of the nucleating bands in the more concentrated sample. 

To conclude, BC appear to be a crucial control parameter that accounts for some of the fluctuations reported earlier on similar systems \cite{Lopez04,Becu04}. The interplay between wall slip and SB formation may have major implications for tuning the flow behavior of complex fluids showing flow instabilities. A full understanding of our experiments still requires a proper combination of the above mentioned theories. The competition between local stress relaxation at the wall via slip and bulk relaxation through SB formation depends on the details of the system, such as the surface treatment and the degree of entanglement in the bulk. A possible microscopic input in the theory could be to mimic surface roughness, i.e. stick BC, by randomizing the alignment of the wormlike micelles at the walls. This is missing from the theoretical work so far. Experiments on a less coarse grained level as was achieved here with USV are also needed to identify micellar orientations at the wall.

\begin{acknowledgments}
The experiments were funded by the SoftComp EU Network of Excellence and performed at CRPP, Pessac, France. The authors thank P.~Ballesta and E.~Laurichesse for technical help and J.~K.~G.~Dhont, S.~Lerouge, and P.~D.~Olmsted for fruitful discussions.
\end{acknowledgments}


\begin{thebibliography}{15}
\bibitem{cells} Our ``smooth'' (``rough'' resp.) walls have a roughness (average standard deviation of height profiles measured from atomic force microscopy) of 15~nm (0.6~$\mu$m resp.).
\bibitem{rheological} Rheological data are collected with a stress controlled rheometer (TA Instruments AR 1000) in Couette cells with inner radius 24~mm, outer radius 25~mm, and height 30~mm. A computer-controlled feedback loop is used to apply quenches between constant shear rates and the corresponding stress response $\sigma(t)$ is recorded.
\bibitem{defs} Velocity profiles are linearly extrapolated to estimate the fluid velocities close to the walls $v(0,t)$ and $v(e,t)$, yielding $\gt(t)=(v(0,t)-v(e,t))/e$. The intersection and the slopes of linear fits of $v(x,t)$ over the two SB respectively yield the position $\delta(t)=e\,\varepsilon(t)$ of the interface between SB and the local shear rates $\gp(t)$ and $\gm(t)$. Due to the USV spatial resolution, this fitting procedure cannot ascertain the presence of SB smaller than 80~$\mu$m, hence the lack of data points with $\varepsilon\lesssim 0.1$ in Fig.~\ref{truerate}(b).
\bibitem{range} The large shear rate range, as compared to Refs.~\cite{Hu05,Miller07}, was achieved by using a special lid that reduces the air-sample interface. In this way instabilities of the free surface at high shear rates due to air inclusions are hindered.
\end{thebibliography}
\end{document}